\documentclass[12pt]{article}
\usepackage{graphicx}

\newcommand{\h}{\hspace*{0.4 cm}}

\newcommand{\cent}{\centerline}
\newcommand{\vs}{\vspace*}

\begin{document}
\thispagestyle{empty}

\linespread{1.0}

.

\vs{2 cm}

\cent{\large\bf MAJORANA, THE NEUTRON, AND THE NEUTRINO:}

\cent{\large\bf SOME ELEMENTARY HISTORICAL MEMORIES $^{(*)}$}

\footnotetext{$^{(*)}$ Article appeared in Hadronic Journal
{\bf 40} (2017) 149. \ Work partially supported by CAPES (Brazil), and by TEC/TNRG (USA)}

\vs{0.3cm}

\begin{center}

{Erasmo RECAMI $^{(**)}$

\footnotetext{$^{(**)}$ Vising Professor c/o UNICAMP with a PVE Fellowship of CAPES/BRASIL.}

\

{\em Facolt\`a di Ingegneria, Universit\`a Statale di Bergamo,
24044--Dalmine (BG), Italy;

INFN--Sezione di Milano, Milan, Italy; \ and

DECOM, Faculty of Electrical Engineering (FEEC), State University at Campinas (UNICAMP), Campinas, Brazil.}}

email: {\bf recami@mi.infn.it}

\end{center}

\vs{1.0cm}

{\bf ABSTRACT} \ -- {\small {At the kind request of the Editor of this Journal, and of TEC/TNRG (Usa), we recall in this article some elementary historical information about the role played by Ettore Majorana[1-17] with regard to the Neutron, and to the Neutrino.
We often do explicitly insert direct quotations from the original written statements --in particular by Edoardo Amaldi[22-24]--, which in some cases, and for some topics, constitute the main available evidence. We also discuss some recent, unjustified rumors about Ettore Majorana and Enrico Fermi.}}

\

{\bf Keywords:} Ettore Majorana; Neutron; Neutrinos; Enrico Fermi; Edoardo Amaldi; Majorana spinors; Majorana algebras; Majorana fermions; Fred Wilczek; History of XX century physics.

\newpage

\section{Introduction}

At the kind request of the Editor of this Journal, and of TEC/TNRG(Usa), we wish to recall some historical elementary information about the role played by Ettore Majorana[1-17] with regard to the Neutron, and the Neutrinos.  Often, we shall explicitly insert direct quotations from the original written statements --in particular by Edoardo Amaldi[22-24]--, which in some cases, and for some topics, constitute the main available evidence. We discuss also some recent unjustified rumors about Ettore Majorana and Enrico Fermi.

\

Let us start with the Neutron.

\section{THE NEUTRON}

As soon as the news of the Joliot-Curie experiments[18] reached Rome
at the very beginning of 1932, Majorana understood that they had discovered the ``neutral proton" without realizing it.  Thus, even before the official announcement of the discovery of the neutron, made just after by Chadwick[19], Majorana was able to explain the structure and the stability of atomic nuclei through protons and neutrons, preceding in this way also the pioneering work of D.Ivanenko[20]: as both E.Segr\'e and E.Amaldi have recounted in detail. \  His colleagues remember that before Easter he had already concluded that protons and neutrons (indistinguishable with respect to the nuclear interaction) were bound by the ``exchange forces" originating from the exchange of their spatial positions only (and not also of their spins, as Heisenberg will instead propose), so as to obtain the alpha
particle (and not the deuteron) as saturated in respect of the binding energy.  Only after that Heisenberg had published his own article on the same argument, Fermi was able to persuade Majorana to meet the famous colleague in Leipzig; and finally Heisenberg does convince Majorana to publish his results in the paper ``{\"U}ber die Kerntheorie"[21]. Majorana's paper on the stability of nuclei was immediately recognized by the scientific community --a rare event, as we know, for his writings-- thanks to that timely ``propaganda" done by
Heisenberg himself.

\h The role of Majorana in connection with the Neutron was mentioned, e.g., in our book[1]; but it is based only on the testimonies by the Colleagues he had in the reasearch group leaded by Enrico Fermi at Rome: in particular, as we were saying, by Edoardo Amaldi[22-24] and by Emilio Segr\'e[25-26]. We therefore prefere to confine ourselves,
here, to quote the original written statements, they constituting the main available evidence.

\subsection{A first written account by E.Amaldi}

A rather extensive account has been written by Amaldi in 1968
in Ref.[23]. We report it here, for historical reasons, even if in Italian ({\em adding our translation of it into English in {\bf Appendix A}}).  We shall report, in the following, a similar statement written by Amaldi in English, and appeared in 1984 in his Ref.[24].

\h {\small { $<<$ Verso la fine di gennaio 1932 cominciarono ad arrivare i fascicoli dei {\em Comptes Rendus} contenenti le classiche note di F.Joliot e
I.Curie sulla radiazione penetrante scoperta da Bothe e Becker. Nella prima di tali note veniva mostrato che la radiazione penetrante, emessa dal berillio sotto l'azione delle particelle alfa emesse dal polonio, poteva trasferire ai protoni, presenti in straterelli di vari materiali idrogenati (come l'acqua o il cellofan), energie cinetiche di circa cinque milioni di elettronvolt. Per interpretare tali osservazioni, i Joliot-Curie avevano in un primo tempo avanzato l'ipotesi che si trattasse di un fenomeno analogo all'effetto Compton… Subito dopo, però, avevano suggerito che l'effetto osservato fosse dovuto a un nuovo tipo di interazione fra raggi gamma e protoni, diversa da quella che interviene nell'effetto Compton.

\h Quando Ettore lesse queste note, disse, scuotendo la testa: Non hanno capito niente: ``probabilmente si tratta di protoni di rinculo prodotti da una particella neutra pesante". Pochi giorni dopo giunse a Roma il fascicolo di {\em Nature} contenente la Lettera all'Editore presentata da J.Chadwick il 17 febbraio 1932 e in cui veniva dimostrata l'esistenza del neutrone sulla base di una classica serie di esperienze…  Subito dopo la scoperta di Chadwick, vari autori compresero che i neutroni dovevano essere uno dei costituenti dei nuclei e cominciarono a proporre vari modelli in cui entravano a far parte particelle alfa, elettroni e neutroni. Il primo a pubblicare che il nucleo \`e costituito soltanto di protoni e neutroni \`e stato probabilmente D.D.Ivanenko… Ma è certo che, prima di Pasqua di quello stesso anno, Ettore Majorana aveva cercato di fare la teoria dei nuclei leggeri ammettendo che i protoni e i neutroni (o ``protoni neutri" come egli diceva allora) ne fossero i soli costituenti e che i primi interagissero con i secondi con forze di scambio delle sole coordinate spaziali (e non degli spin), se si voleva far s\`{\i} che il sistema saturato rispetto all'energia di legame fosse la particella alfa e non il deutone.

\h Aveva parlato di questo abbozzo di teoria agli amici dell'Istituto e Fermi, che ne aveva subito riconosciuto l'interesse, gli aveva consigliato di pubblicare al più presto i suoi risultati, anche se parziali. Ma Ettore non ne volle sapere perch\'e giudicava il suo lavoro incompleto. Allora Fermi, che era stato invitato a partecipare alla Conferenza di Fisica che doveva avere luogo nel luglio di quell'anno a Parigi, nel quadro pi\`u ampio della Quinta Conferenza Internazionale sull'Elettricit\`a, e che aveva scelto come argomento da trattare le propriet\`a del nucleo atomico, chiese a Majorana l'autorizzazione di accennare alle sue idee sulle forze nucleari. Majorana rispose a Fermi che gli proibiva di parlarne o che, se ne voleva proprio parlare, facesse pure ma, in quel caso, dicesse che si trattava di idee di un noto professore di elettrotecnica, il quale fra l'altro doveva essere presente alla Conferenza di Parigi, e che egli, Majorana, considerava come un esempio vivente di come non si dovesse fare la ricerca scientifica.  Fu cos\`{\i} che il 7 luglio Fermi tenne a Parigi il suo rapporto su ``Lo stato attuale della fisica del nucleo atomico" senza accennare a quel tipo di forze che in seguito furono denominate ``forze di Majorana" e che in sostanza erano gi\`a state concepite, sia pure in forma rozza, vari mesi prima.

\h Nel fascicolo della {\em Zeitschrift fuer Physik} datato 19 luglio 1932 apparve il primo lavoro di Heisenberg sulle forze ``di scambio alla Heisenberg", ossia forze che coinvolgono lo scambio delle coordinate sia spaziali che di spin. Questo lavoro suscitò molta impressione nel mondo scientifico: era il primo tentativo di una teoria del nucleo che, per quanto incompleta e imperfetta, permetteva di superare alcune delle difficoltà di principio che fino ad allora erano sembrate insormontabili. Nell'Istituto di Fisica dell'Università di Roma tutti erano oltremodo interessati e pieni di ammirazione per i risultati di Heisenberg, ma al tempo stesso dispiaciuti che Majorana non avesse non dico pubblicato, ma neanche voluto che Fermi parlasse delle sue idee in un congresso internazionale…

\h Fermi si adoper\`o nuovamente perch\'e Majorana pubblicasse qualche cosa, ma ogni suo sforzo e ogni sforzo di noi, suoi amici e colleghi, fu vano. Ettore rispondeva che Heisenberg aveva ormai detto tutto quello che si poteva dire e che, anzi, aveva detto probabilmente {\em anche troppo}. Alla fine per\`o Fermi riusc\`{\i} a convincerlo ad andare all'estero, prima a Lipsia e poi a Copenaghen, e gli fece assegnare dal Consiglio Nazionale delle Ricerche una sovvenzione per tale viaggio che ebbe inizio alla fine di gennaio del 1933 e dur\`o dal 19 gennaio al 5 agosto (1933), a parte una interruzione dal 12 aprile al 5 maggio. $>>$} }

\h For our translation in English of this historical
account, see (as we said above) our {\em Appendix A}.

\

\subsection{From the Amaldi's 1984 {\em Physics Reports} on Neutrons and nuclear fission.}

Let us now quote a different account by E.Amaldi, this time[24] in English, appeared in 1984 in his Ref.[24]

\h {\em {$<<$ ...it appears in order to add a few remarks of historical nature about the contribution of Ettore Majorana to the understanding of nuclear forces. When the issue of the ``Comptes Rendus" containing the first note by F.Joliot and I.Curie[18] on the penetrating radiation discovered by Bothe and Becker arrived at the library of the Institute of Physics of the University of Rome, Ettore read it and, shaking his head said more or less: ``They haven't understood a thing. They probably are observing the recoil protons produced by a heavy neutral particle". A few days later we got in Rome the issue of ``Nature" containing the Letter to the Editor from Chadwick in which he demonstrated the existence of the neutron[19].
In order to understand how Ettore could guess this discovery, which was suggested but certainly not demonstrated by the results of Joliot and Curie, it should be remembered that he was familiar, through a paper published a few years earlier by Giovanni Gentile junior, with the nuclear model proposed in 1927 by Lord Rutherford, already mentioned by
us above. Gentile had shown the inconsistency of that model. The idea, however, that there might exist in nature neutral particles of subatomic dimensions had, as it were, remained in the air, also in Rome. Neither I, nor his other friends, recently questioned, remember whether Ettore Majorana came to the conclusion that the nucleus consists solely of protons and neutrons independently from other authors. What is certain is that before Easter 1932 he worked out a theory of light nuclei, assuming that they consisted solely of protons and neutrons (or `neutral protons', as he then said) and that the former interacted with the latter through exchange forces. He also reached the conclusion that these exchange forces must act only on the space coordinates (and not on the spin) if one wanted the alpha particle, and not the deuteron, to be the system saturated with respect to the binding energy.
He talked about this outline of a theory to his friends at the Institute, and Fermi, who had at once realized its interest, advised him to publish his results as soon as possible, even though they were partial. However, Ettore would not want to know about that, because he considered his work to be incomplete. Thereupon, Fermi, who had been invited to the physics conference which was to take place in July of that year in Paris in the wider framework of the Fifth International Conference on Electricity, and who had chosen as his subject the properties of the atomic nucleus, asked Majorana for permission to mention his ideas on nuclear forces. Majorana forbade Fermi to mention them, but added that if he really must, he should say they were the ideas of a well-known professor of electrical engineering who, incidentally, was to be present at the Paris conference and whom Majorana considered to be a living example of how not to carry out scientific research.
Thus, on July 7, Fermi presented his report in Paris on ``The Present State of the Physics of the Atomic Nucleus" without mentioning the type of force which was subsequently called `Majorana exchange force' and which had actually been thought of, although in a crude form, some months earlier.
The issue of the ``Zeitschrif fuer Physik" dated July 19, 1932 contained Heisenberg's first paper on `Heisenberg exchange forces', namely forces involving the exchange of both the space and spin coordinates.}}

\h {\em {$<<$ Fermi again tried to persuade Majorana to publish something, but all his efforts and those of his friends and colleagues were in vain. Ettore replied that Heisenberg had now said all there was to be said and that, in fact, he had probably even said {\rm too much}. Finally, Fermi succeeded in persuading Majorana to go abroad, first to Leipzig and then to Copenhagen, and obtained a grant from the National Research Council for his journey, which began at the end of January 1933 and lasted from Jan. 19 to Aug. 5 (1933), a part from an interruption
the 12th of April to the 5th of May. Majorana's aversion to publishing or making known in any way his results, which appears from this episode, was part of his general attitude. $>>$}}

\h {\em {$<<$ During the period spent in Leipzig, Majorana became friendly with Heisenberg for whom he always had a great admiration and a feeling of friendship. It was Heisenberg who persuaded him without difficulty, by the sheer weight of his authority, to publish his paper on nuclear theory which appeared the same year[21] in the `Zeitschrift fuer Physik'. $>>$}
}

\

\h Before going on, let us present a portrait of Ettore Majorana in
Fig.1.

\begin{figure}[!h]
\begin{center}
 \scalebox{0.2}{\includegraphics{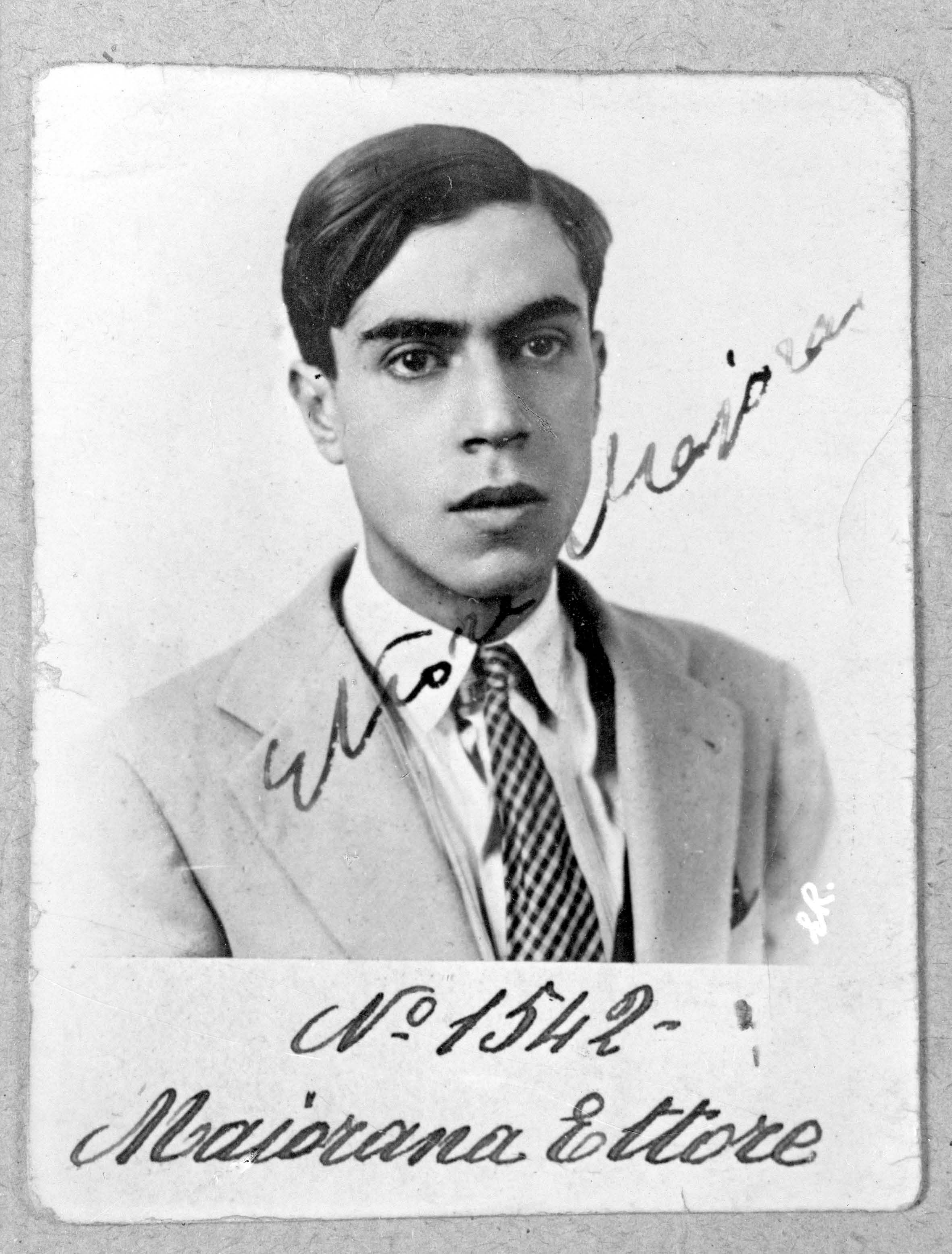}}
\end{center}
\caption{The best portrait of Ettore Majorana, when 23. Reproduction forbidden [copyright by M.Majorana \& E.Recami]. }
\label{Figure1}
\end{figure}

\subsection{On the first attempts to a theory of the Nucleus}

From the same 1984 paper, published by E.Amaldi[23] in ``Physics
Reports", let us report for completeness' sake his history of
{\bf the theory of the nucleus}: Namely, his historical considerations
on {\em the first attempts of a theory of nuclei composed of protons and neutrons only}.

\h {\small {$<<$ Heisenberg, Majorana and Wigner were the first to appreciate fully the importance of the new model describing the nucleus as a system composed only of protons and neutrons. The formalism of non-relativistic quantum mechanics could be applied to it for trying to explain qualitatively, and in part quantitatively, a few nuclear properties. Such a programme clearly involved the introduction in the Schroedinger equation of a ``potential" describing the new forces acting between the nuclear constituents. The choice of this potential was made by the three authors mentioned above, in different ways. Each of them had its grounds and merits and remained as a useful (or even necessary) ingredient of many successive developments. The nuclear forces we consider here were the first example of one of the four fundamental interactions acting among subatomic particles, the strong interaction.
A fascinating presentation of a number of historical aspects of these early stages of nuclear physics can be found in the Proceedings edited
by R.H.Stuewer, in particular in the articles by Wigner and by Peierls.
Everybody was ready to accept that new and very strong forces were required to hold the nucleus together, but very little was known about their behaviour, in particular about their dependence on the distance as well as on other variables. Two important properties, however, were known for a long time: (1) The nuclear forces do not extend to very large distances, beyond the nuclear radius. Any long-range nuclear force strong enough to keep the nuclei together would manifest itself through a modification of molecular behaviour. Hence the nuclear potential should have a ``short-range" , i.e. decrease with the distance between two nuclear constituents faster than the {\em 1/r} Coulomb potential. (2) The binding energy of nuclei rises proportionally to the mass number {\em A} (i.e., the number of its constituents), as it happens, for example, for the energy of the intra-molecular forces of a drop of liquid. This property, known as saturation of the nuclear forces can be obtained by considering various possibilities about their nature, a few of which (actually the simplest) are the following:\hfill\break
(i) Exchange forces;\hfill\break
(ii) Repulsive forces at very short distances (i.e., a ``repulsive   core"),\hfill\break
(iii) Velocity dependent forces;\hfill\break
(iv) Many-body forces.\hfill\break
Only the first possibility was used at the beginning, while for rather weak reasons the others were considered only at successive times.

\h $<<$ In his first paper, dated June 7, 1932, Heisenberg starts from a nucleus composed of protons and neutrons with the additional assumption that the neutron, like the proton, is a particle with spin 1/2, which follows Fermi's statistics.

\h He tries to define the nature of the forces between these particles. On the basis of the analogy with the case of the homopolar forces of chemistry, Heisenberg assumes that between neutrons and protons there is an exchange potential {\em J(r)} due to an exchange of ``negative charge" similar to that responsible for the binding of the molecular hydrogen ion (H$_2^+$). As a comment to this assumption, he writes (in German): ``These exchanges can be made intuitive by means of the picture of an exchange of  ``electrons" without spin and that follow Bose statistics. It is, however, more correct to consider the exchange integral {\em J(r)} as a fundamental property of the neutron-proton pair without trying to reduce it to electron movements".
He also introduces a neutron-neutron attractive interaction {\em K(r)} similar to the exchange potential that gives rise to the H$_2$ molecule, while for the proton-proton pairs he considers only the Coulomb repulsion. By analogy with the molecular case he also assumes {\em J(r)} to be considerably larger than {\em K(r)}.
Thus Heisenberg's Hamiltonian of a nucleus, consisting of {\em Z} protons and {\em N} neutrons, turns out to be the sum of five terms originating from the kinetic energy of the particles, the neutron-proton exchange potential {\em J(r)}, the neutron-neutron exchange potential {\em K(r)}, the proton-proton Coulomb repulsion and the neutron-proton rest energy difference.

\h In writing his Hamiltonian, Heisenberg attributes to each particle, in addition to the usual space and spin coordinates, a fifth variable $\rho^3$ which takes the values +1 for neutrons and -1 for protons and treats it formally as the third component of ``a spin" (later called isotopic spin) with components analogous to the 2 x 2 Pauli spin matrices but acting in a new space that ``has nothing to do with ordinary space".
This formalism was useful (but certainly not necessary) for writing interaction terms that transform a neutron into a proton or vice versa. Only starting from 1936, an increasing number of new phenomena which were gradually discovered and studied in sufficient detail, led to isotopic spin being accepted as a new physical quantity, whose conservation is a generally valid principle in the sense that it is respected in all processes involving strong interactions only.
From a qualitative discussion of his Hamiltonian, Heisenberg deduces a number of interesting consequences. The largest terms in the Hamiltonian are the kinetic energy and the charge-exchange potential {\em J(r)}, both symmetric in neutrons and protons. Therefore, if only these two terms are taken into account, the nuclei with equal number of neutrons and protons are energetically favoured and thus should have the greatest stability. On the other hand the long range Coulomb repulsion between protons and the short range weak attraction {\em K(r)} between neutrons both favour increasing number of neutrons for the stability of heavier nuclei, in agreement with observation.
He then discusses the Schroedinger equation of the recently discovered deuteron, i.e. the neutron-proton system in its ground state. He finds that the corresponding eigenfunction is symmetric in the charge and space coordinates of the two particles which, because of the Pauli principle, should have opposite spin. This prediction is wrong, but at the time of Heisenberg's first paper the spin of the deuteron was not yet known. It was a consequence of Heisenberg's choice of the sign of
$J(r)$ to be the same as for the hydrogen molecular ion and can be cured by changing it. What remains wrong with Heisenberg's forces, irrespective of the sign of $J(r)$, is that the deuteron is the saturated system and not the alpha particle, as it is suggested by the measured values of the mass defect of light nuclei.
The rest of the paper is devoted to the discussion of the properties of intermediate and heavy nuclei, of their binding energy and beta and alpha instability. This leads him to describe, semi-qualitatively, the behaviour of the binding energy of the nuclei as a function of $Z$ and $N$, a subject to which he returns in his second paper, dated July 30, 1932, where he discusses in a form destined to be more or less final, the difference of energy between nuclei with $Z$ and $N$ both even, with $Z$ or $N$ odd, and with $Z$ and $N$ both odd.
It is very interesting to see how he succeeded in arriving at conclusions which (although semi-quantitative) are substantially correct, in spite of the fact that at that time the beta processes with emission of positrons had not yet been observed. They were discovered by Joliot and Curie only at the beginning of 1934.
In the same paper Heisenberg discusses two more problems: the large absorption and scattering of $\gamma$ rays observed in those years by various authors, and the properties of the neutron. It was clearly shown that gamma rays of a few MeV energy are absorbed and scattered in matter, in considerable excess with respect to theoretical estimates based on the Klein-Nishina formula.
The effect was generally assumed to be of nuclear origin and Heisenberg tried to explain it by considering, in addition to the changes of the movements of protons and neutrons inside the nucleus, due to the electric field of the incident radiation, also the ``excitation of the negative charge (electron) bound in each neutron". Because of the small mass of the electron this second mechanism was found to give a contribution large enough for representing the experimental data, and thus provided, in Heisenberg's opinion, an argument in favour of the existence of an electron inside the neutron. In 1933, however, it was found that the ``excess absorption" is due to pair production and the ``excess scattering" to the annihilation of electron-positron
pairs.

\h $<<$ On the other hand if the neutron is regarded as a system composed of one electron and one proton so strongly bound as to have linear dimensions much smaller than the nucleus, one faces, in an aggravated form, the same conceptual difficulties met by any nuclear model with intra-nuclear electrons.
Both the problem of the scattering of gamma rays by nuclei and that of the nature of the neutron are taken up and further investigated by Heisenberg in his third paper dated December 22, 1932. He puts forward two alternative hypotheses on the nature of the neutron: the first is that the neutron is an elementary particle without structure, as assumed by Perrin. However, in order to explain the beta decay of a heavy nucleus it would be necessary for the latter to contain electrons in addition to neutrons and protons. Thus, there arise difficulties, which cast doubts also on the second hypothesis of the structure of the neutron: the neutron consists of an electron and a proton and as such is likely to interact through exchange forces with the proton. The situation appeared to Heisenberg to be so serious as to prompt him to advance the hypothesis that quantum mechanics might be inadequate to describe the phenomena which take place on a nuclear scale.
Today, we know that this is not true and that this difficulty, and others which are not mentioned here, was to disappear with the hypothesis put forward by Yukawa in 1935, that the particles exchanged between any two nucleons, namely protons and neutrons, are not electrons but mesons, namely particles of integral spin (which are, therefore governed by Bose statistics) whose mass is at least 280 times greater than that of the electron. The experimental confirmation of this idea came by steps over a period of 12 years: 1935--1947.
A large part of Heisenberg's third paper is devoted to extending the Thomas-Fermi[27] method to the case of the nucleus, assuming for this purpose a Hamiltonian which differs from that given in the first paper only in so far as, in addition to the exchange forces, there are also ordinary forces acting between protons and neutrons.

\h Heisenberg deduces the expression of the energy of the system in its lowest state, imposing as a condition that the probable value of the Hamiltonian be stationary with respect to arbitrary variations of the density of the particles. At first, the Hamiltonian was calculated by taking for the kinetic energy of the protons and neutrons, as in the Thomas-Fermi method[27], the expression valid for a totally degenerate Fermi gas not subjected to external fields, and a potential energy of the form described above.
Under these conditions, however, it is not possible to avoid the collapse of the nucleus as a result of the fact that the nuclear forces tend to reduce to zero, or at least to an extremely small value, the distance between the two particles of each neutron-proton pair. In order to overcome this drawback Heisenberg had to introduce a cut-off in the potential energy at the shortest distances, a forerunner of what, at later times, was to be called a repulsive core.

\

\h $<<$ Majorana's paper[21], dated Leipzig, March 3, 1933, appeared at this point. After re-examining the hypothesis of the Heisenberg model, Majorana reconsidered the difficulties relating to the structure of the neutron and reached the conclusion that in the current state of knowledge, the only thing to do was to try to establish the law of interaction between proton and neutron on the basis of criteria of simplicity only, but in such a way as to reproduce as correctly as possible the most general and characteristic properties of the nuclei. After having discussed some of these and having observed that the nucleus behaves like a piece of extensive and impenetrable ``nuclear matter" , whose different parts interact only upon immediate contact, and having particularly stressed the fact that the experimental data clearly show a law of proportionality of both the binding energy and the volume with respect to the total number of nucleons present, Majorana introduces an exchange potential acting on the space coordinates alone, whose sign was such as to be attractive in states of even angular momentum. He justifies his choice by observing that by doing this, two important results are obtained. The first is that both neutrons of the a particle exercise an attraction on each proton. The second is that in the approximation in which the Coulomb repulsion is neglected, the eigenfunction of the a particle is totally symmetric in the coordinates of the centres of mass of the four component particles, as it is reasonable to expect since it represents a complete shell.
Turning from the a particle to heavier nuclei, the additional nucleons are forced, because of Pauli's principle, to enter into more excited states and since the exchange energy is high only in the case of particles in the same (or almost the same) orbital state, it is concluded that saturation of both the binding energy and the density is practically reached in the case of the a particles.
The essential differences with respect to Heisenberg's scheme is that the exchange forces concern only the space coordinates and the sign of $J(r)$ used by Majorana is the opposite of that of Heisenberg. Under these conditions the symmetry properties of the eigenfunctions are such that it becomes possible to explain the saturation phenomena without introducing a cut-off of the potential energy. Incidentally, I should recall that in comparing his Hamiltonian to that of Heisenberg, Majorana states that he avoids the ``inconvenient" formalism of the
``$\rho$ spin" by treating the neutron and the proton as different particles.
After these qualitative considerations, Majorana writes the Hamiltonian as the sum of the kinetic energy of all the particles, of the Coulomb repulsion energy between all the proton pairs and of the energy due to the exchange forces between all the proton-neutron pairs. He calculates the corresponding eigenvalue, taking for the eigenfunction of the system the product of two totally antisymmetric eigenfunctions (for the simultaneous exchange of both the space and the spin coordinates), the one relating to the protons and the other to the neutrons. He thus reaches the conclusion that if the Coulomb interaction is neglected, the energy of the system is proportional to the number of particles of which it is composed; in other words he shows quantitatively how, with only Majorana exchange forces, it is possible to obtain the saturation both of the density and the binding energy of the nuclei.

\

\h $<<$ At about the same time of the third paper by Heisenberg, Wigner published a paper on the mass defect of deuteron and helium. In contrast to Heisenberg and Majorana, who used exchange forces, Wigner describes the neutron-proton interaction with an ``ordinary" attractive short-range potential. He knew Heisenberg's papers but ``he disliked exchange forces". Other theoreticians at the time rejected the idea of obtaining saturation by introducing a repulsive core, as it was suggested by the analogy with the inter-atomic forces in molecules or solids. Majorana, for example, noted explicitly: such forces would be aesthetically unattractive.
In his paper Wigner first showed that there must be a sufficiently deep potential well to have any bound state of the proton-neutron system. Therefore the existence of the deuteron gives a definite connection between the width of the proton-neutron potential and the depth of it. Then, by assuming the value of the width of the potential well obtained from the experimental results on the neutron-proton elastic scattering, Wigner showed that it becomes possible to compute the mass defect of other nuclei. Such computations were carried out by Wigner for helium and yielded values greater than the mass defect of deuterium by a large factor, in agreement with experiments.
An important idea contained in Wigner's paper was that the very much greater binding energy per nucleon of the a particle with respect to the deuteron is a consequence of the short-range of the neutron-proton interaction. In the deuteron one of the particles, let us say the neutron, remains out of the potential well for a large fraction of the time so that their binding energy is reduced. This result brought Wigner to propose, a few months later, the idea of zero-range nuclear forces. Such a scheme, however, had to be abandoned when, about two years later, L.H.Thomas showed, ``in a very clever paper", that for a given deuteron binding energy, the binding energy of three nucleons, $^3$H and $^3$He, increases indefinitely as the range is decreased, so that zero range would lead to an infinite binding energy for mass number $A$ = 3.

\

\h $<<$ Thus at the beginning of 1933 the types of nuclear forces proposed by various authors were: (1) ordinary forces or Wigner forces, as they are frequently indicated in the literature; (2) Heisenberg exchange forces; (3) Majorana exchange forces.
A third type of exchange forces was proposed in 1936 which exchanges the spin of the two interacting nucleons leaving their position unaffected. The three operators $P^{\rm H} \; $, $P^{\rm M}$  and  $P^{\rm B}$ that applied to a two nucleon wave function perform the interchange proposed by Heisenberg, Majorana and Barlett, clearly have the following properties: $P^{\rm H}$ = $P^{\rm M}$ $P^{\rm B}$;  with $(P^{\rm H})^2$ = $(P^{\rm M})^2$ = $(P^{\rm B})^2$ =1 .

\h In 1936, however, Yukawa had already proposed the existence of a particle of an intermediate mass and spin zero (the meson) as the mediator of the nuclear forces and the distinction between Wigner forces and exchange forces had become a consequence of the value (te or zero) of the electric charge of the exchanged meson.
In spite of the recognized superiority of the ``meson field" approach, there are many good reasons for using, even today, some ``phenomenological potential", for representing the nucleon-nucleon interaction at low energy ($<<m_{\pi} c^2$ = 140 MeV). In some of the expressions of this type still used today, there appear addictive terms corresponding to Wigner, Heisenberg and Majorana forces.

\

\h $<<$ Let me go back to 1933. By adjusting the potential well of the Majorana forces to be attractive for even values of the angular momentum $\ell$  of the relative motion of a proton and a neutron, the potential becomes repulsive for odd values of
$\ell$. This alternation of the sign of the forces for even and odd $\ell$ is a characteristic feature of exchange forces, and can be used, as pointed out by Wick, for deciding whether the forces acting between neutron and proton are exchange forces or ordinary ones. If the de Broglie wave length is comparable to the range of the forces, it becomes hard to deflect the neutron and small deflections are more likely than large ones. But exchange forces interchange the nature of the particles and the incident neutron, which tends to continue in the forward direction, becomes a proton and the outgoing neutron tends to go backward in the centre of mass frame. $>>$}
}

\section{A few elementary historical notes on the Neutrino}

During the 1920's physicists believed matter was built of electrons and protons only (which they even called ``negative and positive electrons"). They were therefore attempting at explaining the
atomic nucleus as composed of protons and (negative) electrons, basing
themselves also on the observations that protons could be knocked out of light elements by alpha particle bombardment, while electrons emerged in radioactive beta decay (especially from very heavy nuclei).
Majorana himself had tried to work out for the nucleus a consistent
theory of such a type: unsuccessfully, of course --as we know from the
extremely vast amounts of scientific manuscripts[3-7] he left unpublished--. Any other elementary constituent of the atom would have been considered superfluous.

\h But, in 1930, Wolfgang Pauli postulated the existence of the neutrino to explain the continuous distribution of energy of the electrons emitted in beta decay, which seemed to imply a non-conservation of energy in such radioactive decays. Only with the emission of a third particle could momentum and energy be conserved; avoiding moreover apparent spin and statistics anomalies. Indeed, Pauli suggested that a neutral particle of small mass might accompany the electron in nuclear beta decay, calling it (until Chadwick's discovery) the ``neutron", a
name corrected later on into ``neutrino". Indeed, when J.Chadwick discovered a much more massive nuclear particle, in 1932, and also named it a neutron, this left the two particles with the same name.
The name {\em neutrino} was playfully coined by Edoardo Amaldi, to resolve the confusion, during a conversation with Enrico Fermi at the
renowned Institute of Physics in {\em Via Panisperna} street at Rome, as a jocular diminutive of the word neutron. It was in fact a pun on ``neutrone", the Italian equivalent of neutron: the {\em -one} suffix being an augmentative in Italian, so that neutrone could be read as the ``big neutral thing"; while {\em -ino} was replacing the augmentative with a diminutive suffix, the word {\em neutrino} meaning therefore in Italian the ``small neutral thing". Such a term was adopted by Fermi in the Paris Conference of July 1932, and in the 1933 Solvay Conference, where even Pauli accepted and used it.  Then, it spread in the international scientific community, especially when Fermi used it in 1933 in his fundamental theoretical paper on beta decay, famous as being the first theory of weak interactions. In fact, it was the precursor to the subsequent theory in which the interaction between proton–neutron, and electron–antineutrino, is mediated by a virtual
$W$ boson [incidentally, the weak interaction theory, born with Fermi
in 1933, was concluded --in a sense-- with the experimental revelation
at Cern fifty years later, in 1983 of the heavy bosons $W$ and $Z$: that is, of the ``heavy photons"; a discovery attributed mainly to Carlo Rubbia, another Italian, who received his Nobel price
in 1984].

\h Upon the prediction and discovery of a second and third neutrino, it became necessary to distinguish between the different types of neutrinos. Pauli's neutrino is now called the electron neutrino, while the second and third neutrinos is identified as the muon neutrino and
the tau-neutrino, respectively.

\

\subsection{Pauli's letter of the 4th of December 1930}

It is known that Wolfgang Pauli proposed the existence of the
neutrino in a (machine-typed) letter sent by him from Zurich to the physicists who were going to meet in Tuebingen on December 4th, 1930, asking a colleague to take his letter to that Meeting and remain at
the disposal of the audience for any needed further information.

\h Il follows a translation into English of his

\

{\em Open letter
to the group of radioactive people at the Gauverein Meeting in Tubingen}:

\

\noindent Physics Institute\\
of the ETH\\
Zurich

\

{\hfill                             Zurich, Dec. 4, 1930}

{\hfill                             Gloriastrasse}

Dear Radioactive Ladies and Gentlemen,

\h as the bearer of these lines, to whom I graciously ask you to listen, will explain in more detail, prompted by the ``wrong" statistics of the N and Li6 nuclei and the continuous beta spectrum, I had recourse to a  desperate remedy to save the ``exchange theorem" of statistics and the law of conservation of energy. Namely, to the possibility that there could exist in the nuclei electrically neutral particles, that I will tentatively call ``neutrons", which have spin 1/2 and obey the exclusion principle, and which further differ from light quanta in that they do not travel with the velocity of light. The mass of my ``neutrons" should be of the same order of magnitude as the electron mass and in any event not larger than 0.01 proton masses. The continuous beta spectrum would then become understandable by the assumption that in beta decay a ``neutron" is emitted in addition to the electron such that the sum of the energies of the neutron and the electron is constant...

\h I agree that my remedy could seem incredible because one should have seen these ``neutrons" much earlier if they really exist. But only the one who dare can win, and the difficult situation, due to the continuous structure of the beta spectrum, is lighted by a remark of my honoured predecessor, Mr. Debye, who told me recently in Bruxelles: "Oh, it's well better not to think about this at all, like about new taxes". From now on, every solution to the issue must be discussed. Thus, dear radioactive people, look and judge.

\h Unfortunately, I cannot appear in Tubingen personally, since I am indispensable here in Zurich because of a ball on the night of 6/7 December. With my best regards to you, and also to Mr. Back.

Your humble servant,

W. Pauli

\

\

\h Pauli thought his proposal of the ``neutron" was too speculative, and did not publish it in a scientific journal until 1934, by which time Fermi had already developed a theory of beta decay (1933) incorporating the neutrino.

\subsection{Fermi's costruction of the theory of $\beta$-decay and of Weak Intercations}

After Pauli suggested the existence of electrically neutral neutrinos, and Fermi made them the basis --to use the words of F.Wilczek, Nobel laureate-- of an impressive, quantitative theory of beta decay. This
happened[28,29] in 1933, exacly 300 years after the 1633 condamnation
of Galileo and his School.

\h {\small {If I am allowed to insert a digression, let me provide an idea
of what the activities of Fermi and his group meant to Italian physics
and science --and later on for the American ones, as it is more widely known--, by recalling that the Italian physical sciences had once before achieved a position of international superiority --with Galileo.  But while the condemnation on July 22nd, 1633, by the Power existing at that time in Italy (the Church) did not have serious consequences for Galileo himself, it proved disastrous for the School of Galilean physics, which could have continued on as the finest in the world. The vast, promising scientific movement founded by Galileo was cut at the roots by the condemnation of the master, such that physics then totally transferred beyond the Alps.  John Milton, recalling a visit to the the ``famous Galileo, by now old and a prisoner of the Inquisition" (Galileo died in 1642), summed up the situation brilliantly, noting in 1644 that {\em the state of servitude to which science had been reduced in its homeland was the reason why the Italian spirit --so alive before-- was by now extinct, and for many years thereafter everything that was written was nothing more than flattery and platitudes.}  Almost two centuries passed before another great physicist surfaced: Alessandro Volta. Volta created a branch of research that lead to predominantly technological applications by Antonio Pacinotti, Galileo Ferraris and Augusto Righi and, later, to those of Guglielmo Marconi. But this did not yield a true "school" of physics.  So by the end of 1926, when Fermi obtained the chair of theoretical physics in Rome, Italy was certainly not prevalent in the world of physics.
It was only Fermi who, three centuries after Galileo, managed to generate an extensive, modern movement within the physical sciences. For example, the article which initiated Fermi's theory of ``weak interactions" was released in 1933, exactly three hundred years, as
we were saying, after the final sentencing of Galilean theory. This digression into the past might help to clarify the cultural significance, as well as the difficulty, of the re-conquest of Italian physics in the last century. In this context, the presence of Ettore Majorana is potentially crucial. As we know --and shall soon recall--, he was beneath no other theoretical physicist or physico-mathematician.  But, as a pioneer far ahead of his time (and due also to his shyness and self-critical nature), only a few of his articles were quickly understood, appreciated and utilized. Moreover, science lost the man, his work and his leadership very early on. Certainly, if the Italian theoretical physics had been able to draw upon the genius of Majorana (and his students), together with Fermi, for a longer period, the consequences could have been enormous, perhaps unimaginable. This does not change the fact that the fruits of Ettore Majorana's farsighted intellect can be seized --and are more and more seized-- still today - not only in Italy, but all over the world.} }

\

\h Fermi first submitted his ``tentative" theory of beta decay to the famous science journal {\em Nature}, which rejected it ``because it contained speculations too remote from reality to be of interest to the reader".  Nature later regretted that rejection... Fermi then submitted revised versions of the paper to Italian and German journals, which accepted and published them in those languages[28,29] in 1933, and then in 1934. The paper, of course, did not appear at the time in English: An English translation of the seminal paper was published only in 1968 in the {\em American Journal of Physics}.
Fermi found the initial rejection of the paper so troubling that he decided to take some time off from theoretical physics, and do only experimental physics. This had a positive side-effect, leading shortly to his famous work with activation of nuclei by slow neutrons.

\h As we already said, Fermi's theory of beta decay included the neutrino, presumed to be massless as well as chargeless.
Treating the beta decay as a transition that depended upon the strength of the coupling between the initial and final states, Fermi developed a relationship which is now referred to as Fermi's Golden Rule.
Namely, Fermi's Golden Rule says that the transition rate is proportional to the strength of the coupling between the initial and final states, factored by the density of the final states available to the system. But the nature of the interaction which led to beta decay was still unknown in Fermi's time. It took some 20 more years to work out a detailed model of weak interactions which fitted the observations.

\subsection{Just a mention of Majorana neutrino}

While Fermi was constructing his beta-decay teory in 1933, the same year Majorana was working on his symmetric theory of particles
and antiparticles (initiated probably one year before, already),
that he had in mind to apply to neutrinos/antineutrinos.

\h As we have said, from the manuscripts left unpublished by Majorana,
it appears that he was formulating also the essential lines of his symmetrical theory of particles and anti-particles during those years, 1932-1933.  Even though Majorana published that theory years later, on the point of participating in the 1937 competition for professorship: ``Teoria simmetrica dell'elettrone e del positrone"; a publication that was initially noted almost exclusively for having introduced the Majorana representation of the Dirac matrices in a real form.

\h Once Majorana had returned from abroad, where he met several other important
personalities including Bloch, Bohr and Weisskopf, he did not publish any papers for
several years. But his research activity during this period, which focused mainly on
field theory and quantum electrodynamics, is well testified by a number of unpublished
scientific notes, part of which have been reproduced by us in
refs.[5,6] and [4,3]. In 1937, however, probably
after being invited by Fermi to compete for a full professorship, Majorana published what  was to become his most famous paper: ``Symmetric theory of electrons and positrons,"
in which he introduced the so-called Majorana neutrino hypothesis.
A consequence of this theory is that a neutral fermion can coincide
with its anti-particle: This hypothesis was revolutionary because it argued that the antimatter partner of a given matter particle could be the particle itself. And this was in direct contradiction to what Dirac
had successfully assumed in order to solve the problem of negative
energy states in quantum field theory (i.e., the existence of the
positron).

\h But Majorana was just interested --among the others-- in eliminating
the ``Dirac sea" postulate. Let us use the words used by Fermi when
recommending Majorana to the Ministry[1] for attributing to him a
full-professorship independently of the 1937 Competition:

\

$<<$ In a recent paper he devised a brilliant method for treating positive and negative electrons in a symmetric way, finally eliminating the need for the highly artificial and unsatisfactory hypothesis of an infinitely large electric charge diffused throughout space, a question which had been unsuccessfully confronted by numerous other
scholars $>>$.[1]

\

\h Anyway, with unprecedented farsightedness Majorana suggested
that the neutrino, which had just been postulated, as we know,
by Pauli and Fermi,
could be such a particle. This would make the neutrino unique
among the elementary particles and, moreover, enable it to have
mass: a property that favoured the possibility of neutrino
oscillations (a phenomenon, predicted by B.Pontecorvo, and
later on experimentally verified).

\

\h As with Majorana's other writings, this article also started to have luck only decades later, beginning in 1957.  Now expressions like Majorana spinors, Majorana mass, and Majorana neutrinos are are well-known to be highly fashionable. \ As we already
mentioned, Majorana's publications (still little known, despite it all) can be regarded as a goldmine for physics.  For example, it has been
recently  observed by C. Becchi how, in the first pages of that paper, a clear formulation of the quantum action principle appears: the same principle that in following years, for instance through Schwinger's and Symanzik's works, has brought about quite important developments in quantum field theory. \ Moreover, theoreticians[14,31],  and especially
mathematicians, have more recently noticed, and worked out, the quite {\em important} observation that Majorana spinors imply Majorana Algebras and Majorana Involutions, which admit of extraordinary applications, for instance in the study of the famous {\em Monster Group}: See refs.[32,33].

\

\h Let us recall at this point that a very large part of Majorana's work was left unpublished by him.
We have in our hands his Master thesis on  ``The quantum theory of
radioactive nuclei", 5 notebooks ({\em ``Volumetti''}), 18
booklets ({\em ``Quaderni''}), 12 folders with spare papers, and
the set of the lecture notes for the course on theoretical physics
held at the University of Naples. For the interest of E. Amaldi,
these manuscripts were deposited by Luciano Majorana (Ettore's
brother) at the ``Domus Galilaeana" of Pisa, Italy. We already stated that our analysis
of those manuscripts allowed us to ascertain that all the existing
material, except that for the lectures delivered at the Naples
University, seems to have been written by approximately 1933; even
the rough copy of his last article, which Majorana
published in 1937, seems to have been ready by 1933, the year in which the discovery of the positron was confirmed. Indeed, we don't know too much of what he did in the following years, from 1934 to 1938, except for a series of 34 letters written by Majorana between March 17, 1931, and November 16, 1937, in reply to his uncle Quirino ---a renowned experimental physicist and at that time the president of the Italian Physical Society--- who had been pressing Majorana for theoretical
explanations of his own experiments. Such letters, which reveal
once more as Majorana was deeply knowledgeable even about
experimental details, were copied for us in the seventies by Quirino's daughter Silvia Majorana Toniolo, and much later (1989) transmitted by us to G.Dragoni, who ended publishing them (due to their technical nature, in our book[1] we could publish only few of them).  By contrast, his sister Maria recalled that, even in those years,
Majorana, who had reduced his visits to Fermi's Institute,
starting from the beginning of 1934 (that is, after his return
from Leipzig), continued to study and work at home many hours
during the day and at night. What went on he doing. From a letter of his to Quirino, dated January 16, 1936, we find a first answer, because we get to learn that Majorana had been occupied ``since some time, with quantum electrodynamics"; knowing Majorana's love for understatements,
this no doubt means that by 1935 Majorana had profoundly dedicated
himself to original research in the field of quantum
electrodynamics.

\h This seems to be confirmed also by a text written
by Majorana in Frence, and recently retrieved[30], where the author dealt with a peculiar topic in quantum electrodynamics: {\em An interesting topic for us, since he actually dealt again with a hole theory which eliminated the necessity of the the unaesthetic postulate
of the  ``Dirac sea"}. It is instructive, for this, just to quote directly from the Majorana's paper: as we shall do in our Appendix B.

\

\h Let us go back to Wilczek's words[31]. After that Fermi made neutrinos $<<$ the basis of an impressive, quantitative theory of beta decay, it became interesting to reconsider whether one could have spinning particles that are their own antiparticles.
Could one, specifically, have a version of the Dirac
equation that involved real fields? This was a
mathematical question asked, and answered, by Majorana $>>$.

\h The exceptional theory of neutrinos by Majorana would lead us too far, and we are glad to be able to confine ourselves to refer the reader to the excellent presentations of it in works like [14,31], and refs. therein.

\h Let us only stress that, even if the important experiments
presently performed all over the world (e.g., in USA, Japan, Italy,...)
haven't revealed yet if neutrinos are Dirac's or Majorana's, nevertheless, in condended matter physics, structures have been found
several times that are ``Majorana fermions": see Refs.[34-36].

\section{Reply to some recent (unjustified) rumors about Fermi and Majorana}

\subsection{E.Fermi did NOT write down himself the last E.Majorana's article...}

We have been told that rumors arose, e.g. in the USA, about the fact that E.Fermi himself could have written down the last E.Majorana's (1937) article, on neutrinos, on the basis of Majorana's idea. A fact like that is rejected by all people who have been studying
Majorana's writings since decades (we, e.g. discovered E.M.'s
epistulary and documents[1] in the years 1969-1972 --examining
at the same time[3] his scientific papers deposited at the `Domus Galilaeana" of Pisa-- and started to investigate them all about 45 years ago) for the reasons that: (i) the characteristic E.M.'s sharp style appears the same in all his papers, while it is quite different from Fermi's style; \ (ii) E.M. had practically
prepared his 1937 article by 1933, as results from many documents
handwitten by him and to be found e.g. in refs.[6,5], and via the Links, c/o http://www.domusgalilaeana.it/index.php?, to {\em Quaderno} 13, as well as to {\em Quaderno} 17, especially around page 20, and after page 74; \ (iii) Enrico Fermi, even if recognized E.Majorana to be much higher than himself in theoretical physics[1], was a big man, and never would have acted as a Majorana's ``secretary"...!

\h This rumor seems to have been generated by a sentence in ref.[14], due
probably to a wrong translation into English of a phrase contained in
E.Amaldi's memoirs about Majorana. Those reminiscences appeared even as an Appendix to our book[1], because of an Amaldi's kind concession. But let us refer to the version appeared --with an English translation-- as an introduction to volume [17]; and follow M.Casella's[37] examination.
Actually, at page XXIV of [17], 8th paragraph, the second phrase
might me misinterpreted by an unmindful translator. That compounud sentence is made of three clauses, and takes a new (erroneous) meaning
if the subject of the last clause is considered to be the first
subject (Fermi) instead of the second (Majorana), {\em as if the second clause were incidental}: So that the main clause was (wrongly) regarded as made of the first and third... But the first subject is plural (Fermi and friends): hence, it could NOT be linked in any case to the third clause, especially in Italian [namely, in the original Italian phrase, it is obvious that the initial plural subject (Fermi and the various friends) cannot be the subject of the last clause, which refers on the contrary to the second subject (Majorana)].

\linespread{1.0}

\h Let us come to the actual original sentence!: $<<$ Fermi e i vari amici si adoperarono in questo senso, e Majorana infine si convinse a gran fatica a prendere parte al concorso, e mand\`o alla stampa su ``Il Nuovo Cimento" il lavoro sulla teoria simmetrica dell'elettrone e del positrone.$>>$; wherein:  ``Fermi e i vari amici si adoperarono in questo senso" is the first clause; \ ``e" is a conjunction (that could be eliminated, or replaced with ``so that"); \ ``Majorana infine si convinse a gran fatica a prendere parte al concorso" is the second clause; \ ``e" is an ordinary copulative conjunction; \  ``mand\`o alla stampa su ``Il Nuovo  Cimento" il lavoro sulla teoria simmetrica dell'elettrone e del positrone" is the third clause.
In conclusion, in the original Italian, the 2nd and 3rd clauses refer to one and the same subjec (Majorana), without any doubt.

\h The translation of the same passage into English, present in ref.[17] itself, reads: $<<$ The problem naturally was to make Ettore enter the competition, since he did not seem to want to do so, and in any event had not published any physics papers for some years. Fermi and various friends tried to persuade him, and finally Majorana was convinced that he should take part in the examination, and he sent his paper on the Symmetrical theory of electrons and positrons (No. 9) for publication in ``Il Nuovo Cimento".$>>$. According to us, even from this English translation appears that the paper was sent for publication by Majorana!

\h Let us consider another translation, contained in the English version of ref.[1]...: $<<$ Fermi and various friends persevered and finally persuaded Majorana, with great difficulty, to take part in the contest, who thus published his work on the symmetry of the electron and the positron in ``Nuovo Cimento".$>>$  In this case the translation could arise a doubt about the meaning of the {\em who} (Fermi or Majorana?): but our previous considerations, besides the logic of the speech, clarify that {\em who} meant Majorana.

\subsection{E.Fermi went on remaining in GOOD relations with E.Majorana}

Other, less recent rumors, indulge in suggesting that the relationship
between Fermi and Majorana deteriorated, when Majorana --for his own
personal reasons[1]-- stopped going to the Fermi research group, after
having returned from Leipzig.  Somebody recalled that sometimes
Majorana end Fermi discussed very loudly about physics and/or mathematics: But we think such lively discussions to be rather
natural between two strong personalities, who apparently dealt with
each other on an equal footing.

\h We want to show on the contrary, by three documents (Figs.2a,2b, and 3), that, when Majorana disappeared, Fermi was still feeling the highest
friendship and estimation for Ettore.  The first two documents
(Figs.2a,2b) were discovered and first published by this author
(see Ref.[1] and refs. therein), while the third one (Fig.3)
was got and published by E.Amaldi. Their content can be re-published,
just quoting the source; but their anastatic reproduction is restricted by copyrights.

\begin{figure}[!h]
\begin{center}
 \scalebox{0.45}{\includegraphics{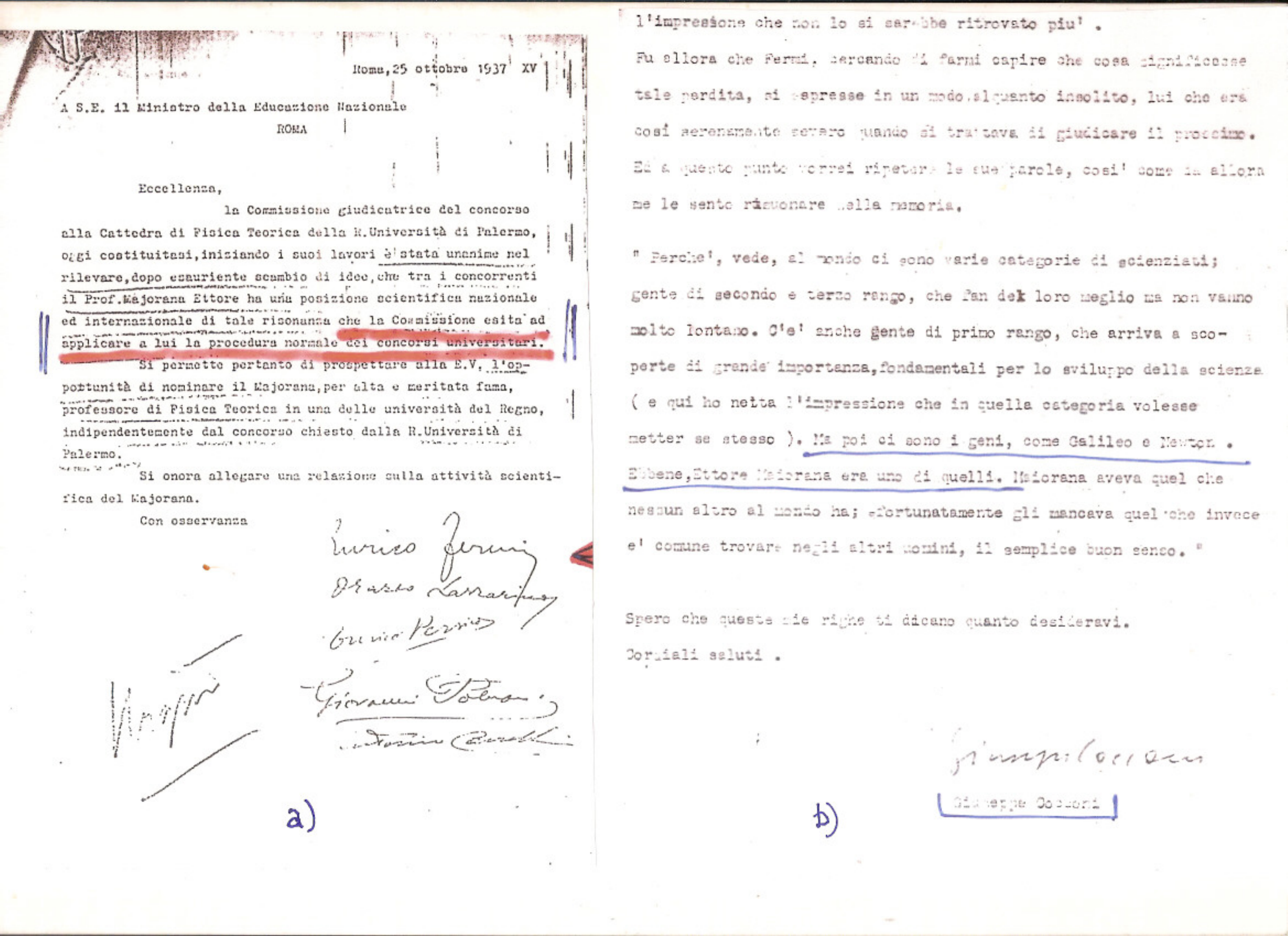}}
\end{center}
\caption{{\bf Fig.2(a)}: Letter to the Ministry of Education on the candidate E.Majorana[1], by the Judging Commission (E.Fermi being the first signer) for the 1937 full-professorship Competition; Rome, Oct. 25th, 1937:  see below for its translation into English. \ 
{\bf Fig.2(b)}:  We reproduce here the principal (final) part of C.Cocconi's testimonial, written at E.Amaldi's request, that recalls how Fermi compared E.Majorana to {\em geniuses like Galileo and Newton}. --  Anastatic reproduction restricted by the relevant copyrights.}
\label{Figures2}
\end{figure}

\

{\bf First document:} -- Letter to the Ministry of Education by the Judging Commission (E.Fermi being the first signer) for the 1937 full-professorship Competition, on the candidate E.Majorana[1].
See {\bf Fig.2(a)}.

\

Our English translation is:

\

Rome, 25 Oct. 1937

To His Excellency The Minister of National Education  -- ROME

Your Excellency,

The Selection Commission to the Chair of Theoretical Physics of the Royal University of Palermo, now constituted, in conducting its operation {\em was unanimous in identifying, after a thorough exchange of opinions, that among the nominees, Prof. Ettore Majorana has a national and international scientific position of such resonance that the Commission hesitates to apply to him the normal procedure for university competitions}.

\h We therefore propose to Your Excellency the opportunity of appointing Majorana, for high and deserved reputation, Professor of Theoretical Physics in a University of the Kingdom, independently of the Competition requested by the Royal University of Palermo.

\h We are honored to attach a report on the scientific activity of Majorana.

\h Respectfully yours,

Enrico Fermi; Orazio Lazzarino; Enrico Persico; Giovanni Polvani; Antonio Carrelli

\

{\bf Second document:} -- Letter of 27 July 1938 by E.Fermi to the Prime Minister Mussolini[1] asking for an increase in the investigations to trace the disappeared E.Majorana [disappeared on March 26, 1938]. It is here reproduced at the {\em last Figure},
{\bf Fig.3}.

\

Enrico Fermi, a 1938 Nobel Prize laureate and one of the greatest physicists of our time (for his accomplishments in 1942 in Chicago perhaps his name will become as legendary as that of Prometheus...), expressed himself in an unusual way on another occasion, when wrote from Rome on July 27th, 1938, to Prime Minister Mussolini asking to intensify the search for Ettore: {\em ``I do not hesitate to declare, and this is not hyperbole, that of all the Italian and foreign scholars whom I had the opportunity to meet, Majorana is the one who for the depth of his genius has impressed me the most.} \
And Bruno Pontecorvo, a direct observer, added: ``A short time after his entry in Fermi's group, Majorana had already acquired so much knowledge and had reached such a level of understanding of physics, that he was able to speak with Fermi about scientific problems as an equal. Fermi himself considered Ettore the greatest theoretical physicist of our time. Often he was left astounded [...]. I remember Fermi's exact words: `If a problem has already been posed, no one in the world can solve it better than Majorana'."
Enrico Fermi was perhaps one of the last examples, and an extraordinary one, of a great theorist who was also a great experimenter. Majorana, instead, was a pure theorist. Indeed (to use the same words as Fermi in the continuation of his letter to Mussolini) Ettore possessed, to the highest degree, that rare combination of skills that make up a theoretical physicist of ``gran classe". Ettore `carried' science, as Sciascia said; indeed he 'carried' theoretical physics. He was not less than a Wigner or Weyl: who, for their aptitude in physics and mathematics, were among the very few scholars for whom Ettore himself harbored unreserved admiration.

\h So, on the one hand, he had no propensity for experimental activities (even if forced, to be clear, he could never have made a tangible contribution to projects like the technological construction of the atomic bomb). But on the other hand, he could descend with unsurpassed and hardly-imaginable depths into the substance of physical phenomena, seeing in them elegant symmetries and powerful new mathematical structures, or uncovering sophisticated physical laws. His sharpness enabled him to see beyond the vision of his colleagues: that is, to be a pioneer. Even his notes, written starting in 1927 when he began his transition from engineering studies to the study of physics, are not only a model of order (they are divided into topics and even have indexes), but also of originality, conciseness, and choice of the essential things only. For this reason these notes, known as the "Volumetti", were suitable for publication just as they were: And indeed they have been published in 2003 by Kluwer Academic Press (in English)[5], and in 2006 by Zanichelli (in Italian)[5]. These ``study" notes are in reality rich with original inventions. Even more so are the remaining manuscripts, which consist of scientific research notes only. But the publication of all these manuscripts is a heavy undertaking; and in 2009 we published with Springer (in English), by another 500 page volume[6], a {\em selection}  of the so-called ``Quaderni": which in their turn contain many, but not all, of the scientific manuscripts left unpublished[7,6] by Majorana.

\begin{figure}[!h]
\begin{center}
 \scalebox{0.4}{\includegraphics{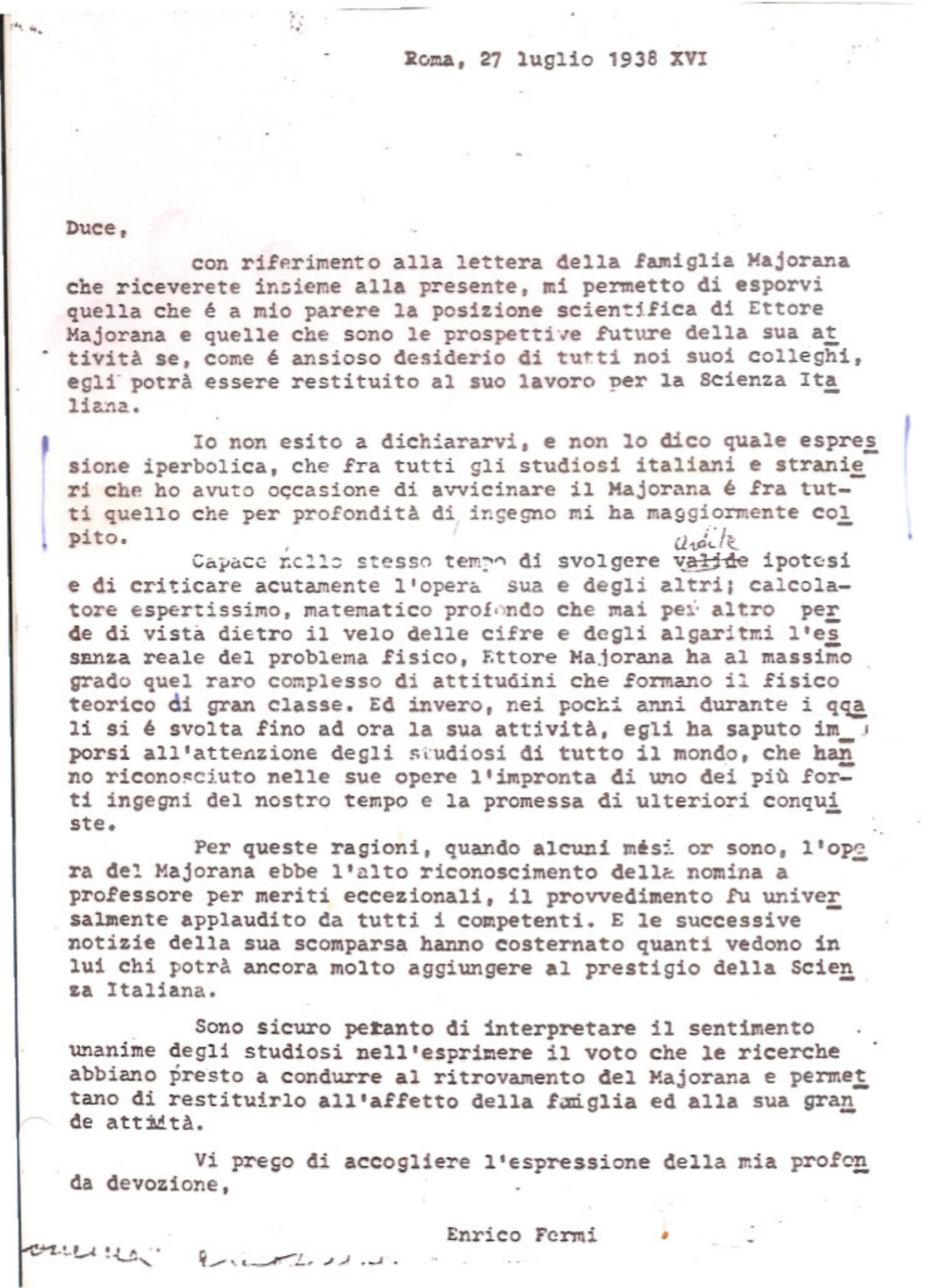}}
\end{center}
\caption{Letter of July 27, 1938, by Enrico Fermi to the Italian Prime Minister Mussolini. It is reproduced {\em in the text}. \ Obs.: at that time photocopiers did not exist, and this is a carbon copy (existing together with a draft).}
\label{Figure3}
\end{figure}

\

{\bf Third document:} -- Letter of July 18, 1965, by G.Cocconi from Geneva: We reproduced in {\bf Fig.2(b)}  the principal (final) part of Cocconi's testimonial, written at E.Amaldi's request, that recalls how Fermi compared E.Majorana to {\em geniuses like Galileo and Newton}.

\h Majorana's fame is firmly justified through testimonials like the one below, which we owe to the mindful pen of Giuseppe Cocconi.  Let us read it in full. From CERN in Geneva, Cocconi (a former collaborator of Enrico Fermi and colleague of Ettore) wrote to Edoardo Amaldi:

$<<$ Geneva, 1965, July 18th - Dear Amaldi, in a discussion that took place long ago on the book [{\em later published by Accademia dei Lincei}] that you are writing about Ettore Majorana, I told you that I, too, had a tenuous contact with Majorana just prior to his disappearance. You expressed then that you wished me to describe my recollections in greater detail, so here I will try to satisfy you.
After having just graduated in January 1938, I was offered, mainly by you, the opportunity to come to Rome for six months as an assistant at the university's Institute of Physics. Once there, I was fortunate enough to join Fermi, G.Bernardini (who had taken a teaching post in Camerino a few months prior) and Ageno (himself also a young graduate), to engage in research regarding the products of the disintegration of ``mu mesons" (then called mesotrons or yukons) produced by cosmic rays. The existence of ``mu mesons" had been proposed about a year earlier, and the problem of their decay was quite contemporary.

$<<$It was indeed while I was with Fermi in the small workshop on the second floor, he working intently at the lathe on a section of a Wilson chamber that was intended to reveal the mesons in the end range, and I busy building a mechanism to illuminate the chamber using the flash produced by the explosion of a strip of aluminum on a shorted battery, that Ettore Majorana came in looking for Fermi. I was introduced to him and we exchanged a few words. A dark face, and that was all. It would have been a quite forgettable episode if, a few weeks later, while with Fermi in the same workshop, I had not heard the news of Majorana's disappearance from Naples. I remember that Fermi busied himself by phoning various places until, after a few days, one got the impression that no one would ever find him.

$<<$ It was then that Fermi, trying to impress upon me the significance of this loss, expressed himself in a rather unusual way for him who was so severe when judging others.  And at this point, I would like to repeat his words just as they echo in my memory: {\em ``Because, you see, in the world there are various categories of scientists. People of second and third rank, who do their best but do not go very far. There are also people of the first rank, who make discoveries of great importance that are fundamental for the development of science (and I have the distinct impression that he would have put himself in this category).  But then there are the geniuses like Galileo and Newton. Well, Ettore was one of these. Majorana had what no one else in the world has. But unfortunately, he lacked what is instead common in other men, plain good sense."}
I hope these lines provide you with what you wished to know.  Kindest regards,\hfill\break
Giuseppe Cocconi.$>>$

\

\noindent ``Plain good sense":  we prefer to say common sense; which may be not always good, or the best.

\section{APPENDIX A}

{\em Let us put here our translation into English of the first detailed account, about the role of E.Majorana in connection with the neutron, written by E.Amaldi in ref.[23], and by us reported above in
its original language (Italian) for historical reasons:}

\

$<<$ Majorana's interest in nuclear physics, which he had already manifested in his thesis, was strongly revived with the appearance of classic works that would lead to the discovery of the neutron at the beginning of 1932. In reality, his renewed interest was part of a new general direction taken by Institute in Via Panisperna, where for a few years there had already been talk of abandoning, albeit gradually, atomic physics, a field in which everyone had worked on for several years, and to focus research efforts principally on the problems of nuclear physics.
Towards the end of January 1932, Comptes Rendus files began arriving with the famous notes of F. Joliot and I. Curie on penetrating radiation discovered by Bothe and Becker. The first of these described how penetrating radiation, emitted from beryllium under the action of alpha particles emitted by polonium, could transfer about five million electron volts of kinetic energy to protons in fine layers of various hydrogenated materials (such as water or cellophane). To interpret these observations, Joliot-Curie had initially suggested that it was a phenomenon analogous to the Compton effect... Soon after, however, they suggested that the observed effect was due to a new type of interaction between gamma rays and protons, different to that which occurs in the Compton effect.
When Ettore read these notes, he said, shaking his head: They did not understand anything, it is probably due to recoiling protons produced by a heavy neutral particle. A few days later, the issue of Nature arrived in Rome containing the Letter to the Editor submitted by J. Chadwick on February 17, 1932, which proved the existence of the neutron on the basis of a classic series of experiments ...
Soon after Chadwick's discovery, several authors realized that neutrons had to be one of the constituents of nuclei and began to offer various models consisting of alpha particles, electrons and neutrons. The first to publish that the core only consisted of protons and neutrons was probably D.D. Ivanenko ... But it is certain that, before Easter of that year, Ettore Majorana had tried to produce a theory for light nuclei assuming that protons and neutrons (or "neutral proton" as they were called then) were the only constituents and that the former interacted with the latter only through exchange forces based on spatial coordinates (and not spin), if you wanted the system saturated with respect to the binding energy to be the alpha particle and not the deuteron.
He mentioned this draft theory to friends at the Institute and to Fermi, who was immediately interested and advised him to publish his results as soon as possible, even if partial. But Ettore was not interested because he judged his work to be incomplete. So Fermi, who had been invited to attend the Conference of Physics that was to take place in July of that year in Paris, in the broader context of the Fifth International Conference on Electricity, and who had chosen the properties of the atomic nucleus as his topic discussion, asked Majorana permission to mention his ideas on nuclear forces. Majorana's reply was to forbid Fermi to discuss them or, or if he absolutely had to, on the condition that he said they were the ideas of a well-known professor of electrical engineering, who was actually going to be present at the Paris Conference, and who Majorana regarded as a living example of how scientific research should not be conducted.
So it was on July 7 in Paris that Fermi gave on his report on "The current status of the physics of the atomic nucleus" without mentioning the kind of forces that were later called "Majorana forces" and that essentially had already been conceived, albeit roughly, several months earlier.
In the Zeitschrift fuer Physik file dated July 19, 1932, appeared Heisenberg's first work on "Heisenberg exchange" forces, i.e., forces that involve the exchange of both spatial and spin coordinates. This work had a large impact on the scientific community: it was the first attempt at a theory of the nucleus that, however incomplete and imperfect, overcame some of the difficulties associated with the principle, which until then had seemed insurmountable. At the Physics Institute of the University of Rome, everyone was extremely interested and full of admiration for Heisenberg's results, but at the same time disappointed that Majorana had not only not published anything, but also did not want Fermi to speak of his ideas in an international congress.
Fermi strove again to convince Majorana to publish something, but every effort of Fermi and his friends and colleagues was in vain. Ettore sustained that Heisenberg had already said everything that could be said on the matter and that, indeed, he had probably said too much. Eventually, however, Fermi was able to convince him to go abroad, first to Leipzig and then to Copenhagen, and he had the National Research Council assign him a grant for the journey, which began at the end of January 1933 and lasted for six or seven months.
His aversion to publish or otherwise reveal his findings seemed, from this episode, to be part of his general attitude. $>>$

\section{APPENDIX B}

As we said in the text, in a letter of his to Quirino, dated January 16, 1936, we learned that Majorana had been occupied ``since some time, with quantum electrodynamics;" knowing Majorana's love for understatements, this no doubt means that by 1935 Majorana had profoundly dedicated himself to original research in the field of quantum electrodynamics.

\h This seems to be confirmed by a recently retrieved text written
by Majorana in French[30], where the author dealt with a peculiar
topic in quantum electrodynamics: {\em Interesting for us since he
dealt with a hole theory which did not imply the unaesthetic postulate of a Dirac sea}.  It is instructive just to quote directly from the Majorana's paper, following the translation presented by S.Esposito in ref.[30]:

\

$<<$ Let us consider a system of $p$ electrons and put the
following assumptions: 1) the interaction between the particles is
sufficiently small allowing to speak about individual quantum
states, so that we may consider that the quantum numbers defining
the configuration of the system are good quantum numbers; 2) any
electron has a number $n > p$ of inner energetic levels, while any
other level has a much greater energy. We deduce that the states
of the system as a whole may be divided into two classes. The
first one is composed of those configurations for which all the
electrons belong to one of the inner states. Instead the second
one is formed by those configurations in which at least one
electron belongs to a higher level not included in the n levels
already mentioned. We will also assume that it is possible, with a
sufficiently degree of approximation, to neglect the interaction
between the states of the two classes. In other words we will
neglect the matrix elements of the energy corresponding to the
coupling of different classes, so that we may consider the motion
of the p particles in the n inner states, as if only these states
exist. Then, our aim is to translate this problem into that of the
motion of $n-p$ particles in the same states, such new particles
representing the holes, according to the Pauli principle.

\

$<<$ Majorana, thus, by following a track due by Heisenberg, applied the formalism of
field quantization to the Dirac's hole theory, obtaining the general expression for the
QED hamiltonian in terms of anticommuting holes quantities. We also point out the
peculiar justification of the use of anticommutators for fermionic variables given by
Majorana; such use, in fact, ``cannot be justified on general grounds, but only by the
particular form of the hamiltonian. In fact, we may verify that the equations of motion
are satisfied to the best by these last exchange relations rather than by the Heisenberg
ones.

$<<$ In the second (and third) part of the same manuscript, Majorana
also considered a reformulation of QED in terms of a photon wave
function, a topic which was particularly studied even in his {\em
Quaderni} (and reported here). Majorana, indeed, reformulated
quantum electrodynamics by introducing a real-valued wave function
for the photon, corresponding only to directly observable degrees
of freedom.

$<<$ In some other notes, perhaps prepared for a seminar at the
University of Naples in 1938, Majorana gave a physical
interpretation of quantum mechanics which anticipated of several
years the Feynman approach in terms of path integral,
independently of the underlying mathematical formulation. The
starting point in Majorana's paper was to search for a meaningful
and clear formulation of the concept of quantum state. Then, the
crucial point in the Feynman formulation of quantum mechanics,
namely that of considering not only the paths corresponding to
classical trajectories, but all the possible paths joining an
initial point with the end one, was introduced after a discussion
on an interesting example of the harmonic oscillator. We also
stress the key role played by the symmetry properties of the
physical system in the Majorana analysis; a feature which is quite
common in papers of this author. $>>$

\

\

{\bf Acknowledgements}\\

The author thanks the Main Editor of this Journal
and TEC/TNRG, Usa, for a kind invitation. He is visiting, as a  ``Bolsista CAPES/BRASIL", the DECOM/FEEC of the State University of Campinas, SP, Brazil. \ The author acknowledges useful discussions  and kind collaboration with Hugo E. Hernandez Figueroa, M. Zamboni Rached, and P.Cardieri, Walmir de Freitas F., K. McDonald, P.L. Dias Peres. \ Thanks are also due to C. Giardini, C. Meroni, S. Paleari, V. Re, P. Riva and C. Rizzi for their kind interest.

\

\

\

\section{REFERENCES}

[1] E.Recami: {\em Il Caso Majorana: Epistolario, Documenti, Testimonianze}, first editions  Mondadori, Milan (1987) and Oscar Mondadori (1991); further editions having been published by
Di Renzo Editore, Rome [{\em www.direnzo.it }] in 2000, 2002, 2008, 2011.

[2] E.Recami: see, e.g., ``Ettore Majorana: The Scientist and the Man", {\em Int. J. Mod. Phys.} {\bf 23} (2014) 1444009 [17 pages], and references therein.

[3] R.Mignani, M.Baldo and E.Recami: ``About a Dirac--like equation for the photon, according to Ettore Majorana," {\em Lett. Nuovo Cim.} {\bf 11} (1974) 568.

[4] {\em {Ettore Majorana -- Lezioni all'Universit\`a di Napoli}}, ed. by B.Preziosi et al. (Bibliopolis pub.; Naples, 1987): book of 199 pages, containing the {\em anastatic} reproduction of the original (initial ten) notes handwritten by Majorana for the lectures he delivered at the beginning of 1938 at Naples University. \ The complete set of the 16 lecture notes (including the Moreno Document) has been typewritten in {\em Ettore Majorana -- Lezioni di Fisica Teorica}, edited by S. Esposito (Bibliopolis pub.; Naples, 2006).

[5] S. Esposito, E. Majorana Jr., A. van der Merwe and E. Recami: {\em Ettore Majorana - Notes on Theoretical Physics} (Kluwer Acad. Pubs.; Dordrecht, 2003); book of 512 pages. \ The original Italian version, instead, has been published in the book: S. Esposito and E. Recami: {\em Ettore Majorana -- Appunti inediti di Fisica
teorica} (Zanichelli pub.; Bologna, 2006).

[6] S.Esposito, E.Recami, A. van der Merwe and R.Battiston: {\em
E.Majorana -- Unpublished Research Notes on Theoretical
Physics} (Springer; Berlin, 2009); book of 487 pages.

[7] M.Baldo, R.Mignani e E.Recami: ``Catalogo dei manoscritti scientifici
inediti di E.Majorana", in {\em E. Majorana -- Lezioni all'Universit\`a di
Napoli} (Bibliopolis pub.; Napoli, 1987), p.175. \ See also Ref.[3].

[8] E.Giannetto: ``Su alcuni manoscritti inediti di E. Majorana" {\em Atti del IX Congresso Nazionale di Storia della Fisica}, ed. by
F.Bevilacqua (Milan, 1988) 173.

[9] D.Fradkin: ``Comments on a paper by Majorana concerning elementary particles,"  {\em Am. J. Phys.} {\bf 34} (1966) 314. \ Cf. also R.Casalbuoni: ``Majorana and the infinite component wave equations",
arXiv: hep-th/0610252.

[10] R.Penrose: ``Newton, quantum theory and reality", in {\em 300 Years of Gravity}, ed. by S.W. Hawking \& W. Israel (Cambridge Univ.Press; Cambridge, 1987); J. Zimba e R. Penrose: {\em Stud. Hist. Phil. Sci.} {\bf 24} (1993) 697; R. Penrose: {Ombre della Mente (Shadows of the Mind)} (Rizzoli; 1996), pp.338-343 and 371-375.

[11] C.Leonardi, F.Lillo, A. Vaglica and G. Vetri: ``Majorana and
Fano alternatives to the Hilbert space,'' in {\em Mysteries, Puzzles, and Paradoxes in Quantum Mechanics}, ed. by R.Bonifacio (A.I.P.; Woodbury, N.Y., 1999), p.312.

[12] {\em Majorana Legacy in Contemporary Physics},
ed. by I.Licata (Di Renzo Pub.; Rome, 2006); this book appeared also in electronic form in {\em Electron. J. Theor. Phys.} {\bf 3} (2006), issue no.10.

[13] E.Arimondo, C.Clark and W.Martin: ``Ettore Majorana and the birth of autoionization", {\em Rev. Mod. Phys.} {\bf 82} (2010) 1947.

[14] F.Wilczek: ``Majorana returns", {\em Nature Physics} {\bf 5} (2009) 614-618.

[15] A.A.Ivanov: {\em Monster Group and Majorana Involutions}, Cambridge Univ. Press (2009).

[16] S.Esposito: {\em The Physics of Ettore Majorana} Cambride Univ.
Press (2014).

[17] {\em Ettore Majorana - Scientific Publications}, ed. by G.F.Bassani and the S.I.F. council (printed by S.I.F., Bologna, and Springer, Berlin; 2006).

[18] I.Curie and F.Joliot: ``The emissipon of high energy photons from hydrogenous substances irradiated
with very penetrating Alpha rays", {\em C. R. Acad. Sci. Paris} {\bf 194} (1932) 273.

[19] J.Chadwick:   ``Possible existence of a Neutron", {\em Nature} {\bf 192} (1932) 312; and ``The existence of the neutron", {\em Proc. Roy. Soc.} {\bf A136} (1932) 692.

[20] D.Iwanenko: ``The neutron hypothesis", {\em Nature} {\bf 129} (1932) 798.

[21] E.Majorana: ``Ueber die Kerntheorie", {Z. f. Phys.} {\bf 82} (1933) 137-145.

[22] E. Amaldi {\em La vita e l'opera di E. Majorana} (Accademia dei Lincei, Roma, 1966).

[23] E. Amaldi: ``Ricordo di Ettore Majorana", {\em Giornale di Fisica}{\bf 9} (1968) 300.

[24] E. Amaldi: ``From the discovery of the neutron to the discovery of
nuclear fission", {\em Physics Reports} {\bf 111} (1984) 1-322.

[25] E. Segré: {\em Enrico Fermi, Physicist} (Chicago Univ. Press, 1970).

[26] E. Segré: {\em Autobiografia di un Fisico} (Il Mulino, 1995); and {\em A mind always in motion} (University of California Press;
Berkeley, 1993).

[27] S.Esposito: ``Majorana solution of the Thomas-Fermi equation", {\em Am. J. Phys.} {\bf 70} (2002) 852; \ ``Majorana transformation for differential equations," {\em Int. J. Theor. Phys.} {\bf 41} (2002) 2417; \ E.Di Grezia and S.Esposito: ``Fermi, Majorana and the statistical model of atoms,'' {\em Found. Phys.} {\bf 34} (2004) 1431.

[28] E.Fermi: ``Tentativo di una teoria dei raggi $\beta$" {La Ricerca
Scientifica} {\bf 4} (1933) 491-495 (in Italian) [also quoted as
Year {\bf 4th}, vol.2, issue 12].

[29] E.Fermi: ``Versuch einer Theorie der beta-Strahlen - I," {\em Zeitschrift fuer Physik} {\bf 88} (1934) 161-177 (in German).

[30] S.Esposito: ``Hole theory and Quantum Electrodynamics in an unknown manuscript in French by Ettore Majorana" {\em Found. Phys.}
{\bf 37} (2007) 956.  See also \ ``A peculiar lecture by Ettore Majorana" {\em Eur. J. Phys.} {\bf 27} (2006) 1147; \ ``Majorana and the path-integral approach to Quantum Mechanics" {\em Ann. Fond. Louis de Broglie} {\bf 31} (2006) 1; \ E.Di Grezia and S.Esposito: ``Majorana and the quasi-stationary states in Nuclear Physics", arXiv:physics/0702179; \ A.Drago and S.Esposito: ``Following Weyl on Quantum Mechanics: the contribution of Ettore Majorana", {\em Found. Phys.} {\bf 34} (2004) 871.

[31] F.Wilczek: lecture notes at MIT, e.g., on ``Majorana-ism" and ``Majorana algebra" (2013),  see:

http://web.mit.edu/8.701/www/Lecture

ism+``algebraFA13.pdf .

[32] A.A.Ivanov: {\em The Monster Group and Majorana Involutions}
(Cambridge Univ. Press, 2009).

[33] M.Whybrow: ``What is Majorana Theory?" (2016), see

https://madeleinewhybrow.files.wordpress.com/2016/01/majorana-

website.pdf .

See also A.Castllo Ramirez: ``On Majorana Algebras and
Representations", PhD Thesis, supervised by A.Ivanov (Imperial College, London, 2014).

[34] R.F.Service: ``Search for Majorana Fermions in Superconductors", {\em Science}{\bf 332}(2011) 193-195; \ C.W.J.Beenakker:  ``Search for Majorana Fermiomns in Superconconductors", {\em Ann. Rev. Cond. Matt. Phys.} {\bf 4} (2013) 113-136.

[35] V.Mourik et al.: ``Signature of Majorana Fermions in Hybrid Superconductor-Semiconductor nanowire devices", {\em Science} {\bf 336}
(2012) 1003-1007; \ A.Das et al.: ``Zero-bias peaks and splitting in a nanowire topological superconductor as a signature of Majorana Fermions", {\em Nature Physics} {\bf 8} (2012) 887-895.

[36] S.Nadj-Perge et Al.:  ``Observation of Majorana Fermions in ferromagnetic atomic chain on a superconductor", {\em Science}
{\bf 346} (2014) 602-607; with a 21 pages comment published online in
{\em Sciencemag} on April 16, 2016:

http://science.sciencemag.org/content/346/6209/602.full .

[37] M.Casella: {\em private communications}.

\end{document}